# Dynamical reentrance and geometry imposed quantization effects in Nb–AlO$_x$–Nb Josephson junction arrays

Fernando M. Araújo-Moreira[1] and Sergei Sergeenkov[2]

[1]Grupo de Materiais e Dispositivos, Centro Multidisciplinar para o Desenvolvimento de Materiais Cerâmicos, Departamento de Física, Universidade Federal de São Carlos, 13565-905 São Carlos, SP, Brazil
[2]Departamento de Física, CCEN, Universidade Federal da Paraíba, Cidade Universitária, 58051-970 João Pessoa, PB, Brazil

**Abstract**. In this review, we report on different phenomena related to the magnetic properties of artificially prepared highly ordered (periodic) two-dimensional Josephson junction arrays (2D-JJA) of both shunted and unshunted Nb–AlO$_x$–Nb tunnel junctions. By employing mutual-inductance measurements and using a high-sensitive home-made bridge, we have thoroughly investigated (both experimentally and theoretically) the temperature and magnetic field dependence of complex AC susceptibility of 2D-JJA.
After brief description of the measurements technique and numerical simulations method, we proceed to demonstrate that the observed dynamic reentrance (DR) phenomenon is directly linked to the value of the Stewart-McCumber parameter $\beta_C$. By simultaneously varying the inductance related parameter $\beta_L$, we obtain a phase diagram $\beta_C$-$\beta_L$ (which demarcates the border between the reentrant and non-reentrant behavior) and show that only arrays with sufficiently large value of $\beta_C$ will exhibit the DR behavior.
The second topic of this review is related to the step-like structure (with the number of steps $n = 4$ corresponding to the number of flux quanta that can be screened by the maximum critical current of the junctions) which has been observed in the temperature dependence of AC susceptibility in our unshunted 2D-JJA with $\beta_L(4.2K) = 30$ and attributed to the geometric properties of the array. The steps are predicted to manifest themselves in arrays with $\beta_L(T)$ matching a "*quantization*" condition $\beta_L(0)=2\pi(n+1)$.
In conclusion, we demonstrate the use of the scanning SQUID microscope for imaging the local flux distribution within our unshunted arrays.

## 1. Introduction

Many unusual and still not completely understood magnetic properties of Josephson junctions (JJs) and their arrays (JJAs) continue to attract attention of both theoreticians and experimentalists alike (for recent reviews on the subject see, e.g. Newrock et al 2000, Araujo-Moreira et al 2002, Li 2003, Kirtley et al 1998, Altshuler and Johansen 2004 and further references therein). In particular, among the numerous spectacular phenomena recently discussed and observed in JJAs we would like to mention the dynamic temperature reentrance of AC susceptibility (Araujo-Moreira et al 2002) closely related to paramagnetic Meissner effect (Li 2003) and avalanche-like magnetic field behavior of magnetization (Altshuler and Johansen 2004, Ishikaev et al 2000). More specifically, using highly sensitive SQUID magnetometer, magnetic field jumps in the magnetization curves associated with the entry and exit of avalanches of tens and hundreds of fluxons were clearly seen in SIS-type arrays (Ishikaev et al 2000). Besides, it was shown that the probability distribution of these processes is in good agreement with the theory of self-organized criticality (Jensen 1998). It is also worth mentioning the recently observed geometric quantization (Sergeenkov and Araujo-Moreira 2004) and flux induced oscillations of heat capacity (Bourgeois et al 2005) in artificially prepared JJAs as well as recently predicted flux driven temperature oscillations of thermal expansion coefficient (Sergeenkov et al 2007) both in JJs and JJAs. At the same time, successful adaptation of the so-called two-coil mutual-inductance technique to impedance measurements in JJAs provided a high-precision tool for investigation of the numerous magnetoinductance (MI) related effects in Josephson networks (Martinoli and Leeman 2000, Meyer et al 2002, Korshunov 2003, Tesei et al 2006). To give just a few recent examples, suffice it to mention the MI measurements (Meyer et al 2002) on periodically repeated Sierpinski gaskets which have clearly demonstrated the appearance of fractal and Euclidean regimes for non-integer values of the frustration parameter, and theoretical predictions (Korshunov 2003) regarding a field-dependent correction to the sheet inductance of the proximity JJA with frozen vortex diffusion. Besides, recently (Tesei et al 2006) AC magnetoimpedance measurements performed on proximity-effect coupled JJA on a dice lattice revealed unconventional behaviour resulting from the interplay between the frustration f created by the applied magnetic field and the particular geometry of the system. While the inverse MI exhibited prominent peaks at $f = 1/3$ and at $f = 1/6$ (and weaker structures at $f = 1/9, 1/12, ..$ ) reflecting vortex states with a high degree of superconducting phase coherence, the deep minimum at $f = 1/2$ points to a state in which the phase coherence is strongly suppressed. More recently, it was realized that JJAs can be also used as quantum channels to transfer quantum information between distant sites (Ioffe et al 2002, Born et al 2004, Zorin 2004) through the implementation of the so-called superconducting qubits which take advantage of both charge and phase degrees of freedom (see, e.g., Krive et al 2004 and Makhlin et al 2001 for reviews on quantum-state engineering with Josephson-junction devices).

Artificially prepared two-dimensional Josephson junctions arrays (2D-JJA) consist of highly ordered superconducting islands arranged on a symmetrical lattice coupled by Josephson junctions (figure **1**), where it is possible to introduce a controlled degree of disorder. In this case, a 2D-JJA can be considered as the limiting case of an extreme inhomogeneous type-II superconductor, allowing its study in samples where the disorder is nearly exactly known. Since 2D-JJA are artificial, they can be very well characterized. Their discrete nature, together with the very well-known physics of the Josephson junctions, allows the numerical simulation of their behavior (see very interesting reviews by Newrock *et al* 2000 and by Martinoli *et al* 2000 on the physical properties of 2D-JJA).

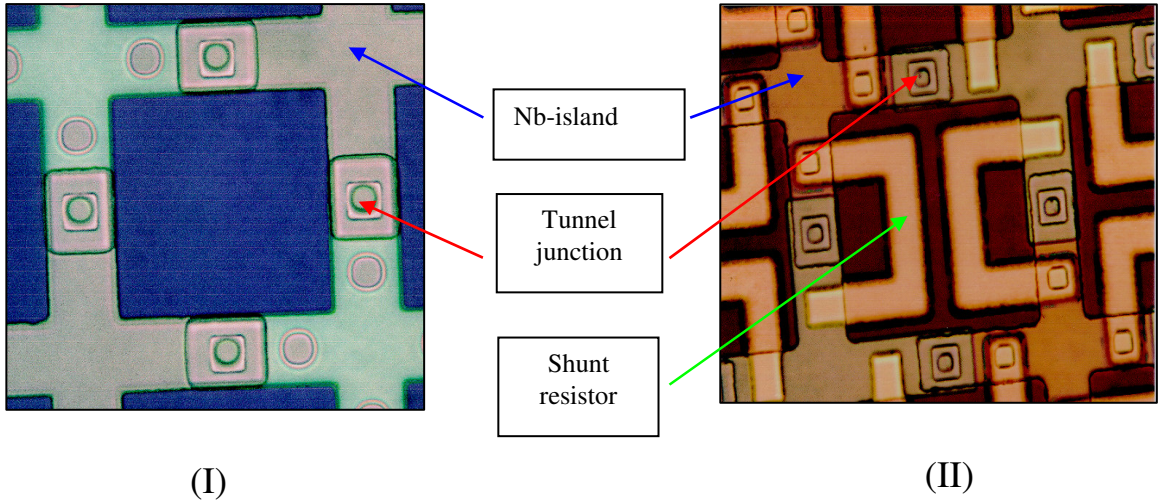

(I)    (II)

**Figure 1.** Photograph of unshunted (I) and shunted (II) Josephson junction arrays.

Many authors have used a parallelism between the magnetic properties of 2D-JJA and granular high-temperature superconductors (HTS) to study some controversial features of HTS. It has been shown that granular superconductors can be considered as a collection of superconducting grains embedded in a weakly superconducting - or even normal - matrix. For this reason, granularity is a term specially related to HTS, where magnetic and transport properties of these materials are usually manifested by a two-component response. In this scenario, the first component represents the *intragranular* contribution, associated to the grains exhibiting ordinary superconducting properties, and the second one, which is originated from *intergranular* material, is associated to the weak-link structure, thus, to the Josephson junctions network (Clark 1968, Saxena *et al* 1974, Yu and Saxena 1975, Resnick *et al* 1981, Sergeenkov 2001, Sergeenkov 2006, Sergeenkov and Araujo-Moreira 2004, Sergeenkov *et al* 2007). For single-crystals and other nearly-perfect structures, granularity is a more subtle feature that can be envisaged as the result of a symmetry breaking. Thus, one might have granularity on the nanometric scale, generated by localized defects like impurities, oxygen deficiency, vacancies, atomic substitutions and the genuinely *intrinsic* granularity associated with the layered structure of perovskites. On the micrometric scale, granularity results from the existence of extended defects, such as grain and twin boundaries. From this picture, granularity could have many contributions, each one with a different volume fraction (Araujo-Moreira *et al* 1994, Araujo-Moreira *et al* 1996, Araujo-Moreira *et al* 1999, Passos *et al* 2000). The small coherence length of HTS implies that any imperfection may contribute to both the weak-link properties and the flux pinning. This leads to many interesting peculiarities and anomalies, many of which have been tentatively explained over the years in terms of the granular character of HTS materials. One of the controversial features of HTS elucidated by studying the magnetic properties of 2D-JJA is the so-called Paramagnetic Meissner Effect (PME), also known as Wohlleben Effect. In this case, one considers first the magnetic response of a granular superconductor submitted to either an AC or DC field of small magnitude. This field should be weak enough to guarantee that the critical current of the intergranular material is not exceeded at low temperatures. After a zero-field cooling (ZFC) process which consists in cooling the sample from above its critical temperature ($T_C$) with no applied magnetic field, the magnetic response to the application of a magnetic field is that of a perfect diamagnet. In this case, the intragranular screening currents prevent the magnetic field from entering the grains, whereas intergranular currents flow across the sample to ensure a null magnetic flux throughout the whole specimen. This temperature dependence of the magnetic response gives rise to the well-known double-plateau behavior of the DC susceptibility and the corresponding double-drop/double-peak of the complex AC magnetic susceptibility (Araujo-Moreira *et al* 1994, Araujo-Moreira *et al* 1996, Araujo-Moreira *et al* 1999, Passos *et al* 2000, Goldfarb *et al* 1992). On the other hand, by cooling the sample in the presence of a magnetic field, by following a field-cooling (FC) process, the screening currents are restricted to the intragranular contribution (a situation that remains until the temperature reaches a specific value below which the critical current associated to the intragrain component is no longer equal to zero). It has been experimentally confirmed that intergranular currents may contribute to a magnetic behavior that can be either paramagnetic or diamagnetic. Specifically, where the intergranular magnetic behavior is paramagnetic, the resulting magnetic susceptibility shows a striking reentrant behavior. All these possibilities about the signal and magnitude of the magnetic susceptibility have been extensively reported in the literature, involving both LTS and HTS materials (Wohlleben *et al* 1991, Braunich *et al* 1992, Kostic *et al* 1996, Geim *et al* 1998). The reentrant behavior mentioned before is one of the typical signatures of PME. We have reported its occurrence as a reentrance in the temperature behavior of the AC magnetic susceptibility of 2D-JJA (Araujo-Moreira *et al* 1997, Barbara *et al* 1999). Thus, by studying 2D-JJA, we were able to demonstrate that the appearance of PME is simply related to trapped flux and has nothing to do with manifestation of any sophisticated mechanisms, like the presence of pi-junctions or unconventional pairing symmetry. To perform this work, we have used numerical simulations and both the mutual-inductance and the scanning SQUID microscope experimental techniques.

The paper is organized as follows. In Section 2 we briefly outline the main concepts related to the mutual-inductance technique. In Section 3 we review the theoretical background for numerical simulations based on a unit cell containing four Josephson junctions. In Section 4 we study the origin of dynamic reentrance and discuss the role of the Stewart-McCumber parameter in the observability of this phenomenon. In Section 5 we present the manifestation of completely novel geometric effects recently observed in the temperature behavior of AC magnetic response. In Section 6 we demonstrate the use of scanning SQUID microscope for imaging the local flux distribution within our unshunted arrays. And finally, in Section 7 we summarize the main results of this work.

## 2. The mutual-inductance technique

Complex AC magnetic susceptibility is a powerful low-field technique to determine the magnetic response of many systems, like granular superconductors and Josephson junction arrays. It has been successfully used to measure several parameters such as critical temperature, critical current density and penetration depth in superconductors. To measure samples in the shape of thin films, the so-called *screening method* has been developed. It involves the use of primary and secondary coils, with diameters smaller than the dimension of the sample. When these coils are located near the surface of the film, the response, i.e., the complex output voltage *V*, does not depend on the radius of the film or its properties near the edges. In the reflection technique (Jeanneret *et al* 1989) an excitation coil (primary) coaxially surrounds a pair of counter-wound pick up coils (secondaries). When there is no sample in the system, the net output from these secondary coils is close to zero since the pick up coils are close to identical in shape but are wound in opposite directions. The sample is positioned as close as possible to the set of coils, to maximize the induced signal on the pick up coils (figure 2).

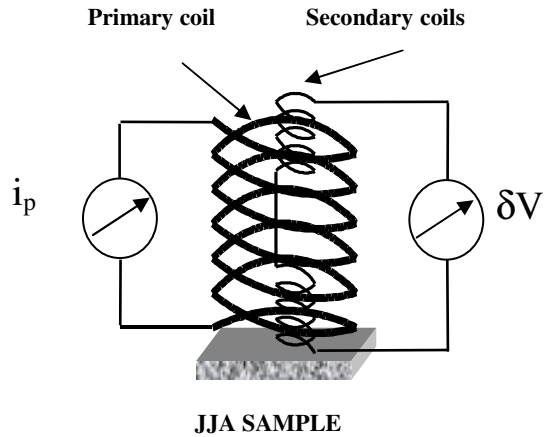

**Figure 2.** Screening method in the reflection technique, where an excitation coil (primary) coaxially surrounds a pair of counter-wound pick up coils (secondaries).

An alternate current sufficient to create a magnetic field of amplitude $h_{AC}$ and frequency *f* is applied to the primary coil. The output voltage of the secondary coils, *V*, is a function of the complex susceptibility, $\chi_{AC} = \chi' + i\chi''$, and is measured through the usual lock-in technique. If we take the current on the primary as a reference, *V* can be expressed by two orthogonal components. The first one is the inductive component, $V_L$ (in phase with the time-derivative of the reference current) and the second one the quadrature resistive component, $V_R$ (in phase with the reference current). This means that $V_L$ and $V_R$ are correlated with the average magnetic moment and the energy losses of the sample, respectively.

We used the screening method in the reflection configuration to measure $\chi_{AC}(T)$ of Josephson junction arrays. Measurements were performed as a function of the temperature T (1.5K < T < 15K), the amplitude of the excitation field $h_{AC}$ (1 mOe < $h_{AC}$ < 10 Oe), and the external magnetic field $H_{DC}$ (0 < $H_{DC}$ < 100 Oe) parallel with the plane of the sample (figure 3).

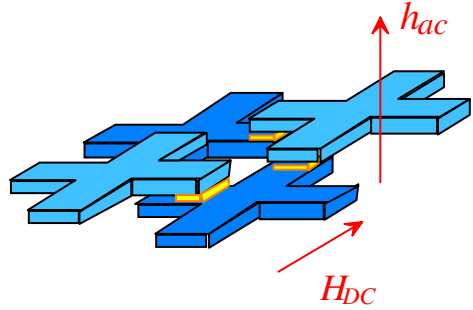

**Figure 3.** Sketch of the experimental setup, where the excitation field $h_{ac}$ and the external magnetic field $H_{dc}$ are respectively perpendicular and parallel to the plane of the sample.

## 3. Numerical simulations: theoretical background

We have found that all the experimental results obtained from the magnetic properties of 2D-JJA can be qualitatively explained by analyzing the dynamics of a single unit cell in the array (Araujo-Moreira *et al* 1997, Barbara *et al* 1999).

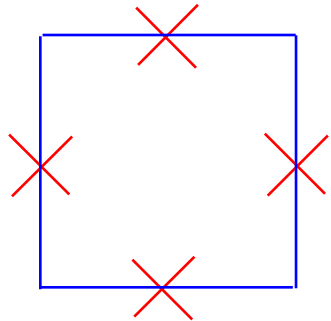

**Figure 4.** Unit cell of the array, containing a loop with four identical junctions.

In our numerical simulations, we model a single unit cell as having four identical junctions (see figure 4), each with capacitance $C_J$, quasi-particle resistance $R_J$ and critical current $I_C$. If we apply an external field of the form:

$$H_{ext} = h_{AC} \cos(\omega t) \quad (3.1)$$

then the total magnetic flux, $\Phi_{TOT}$, threading the four-junction superconducting loop is given by:

$$\Phi_{TOT} = \Phi_{EXT} + LI \quad (3.2)$$

where $\Phi_{EXT} = \mu_0 a^2 H_{EXT}$ is the flux related to the applied magnetic field with $\mu_0$ being the vacuum permeability, I is the circulating current in the loop, and L is the inductance of the loop. Therefore the total current is given by:

$$I(t) = I_C(T) \sin \phi_i(t) + \frac{\Phi_0}{2\pi R_j} \frac{d\phi_i}{dt} + \frac{C_j \Phi_0}{2\pi} \frac{d^2 \phi_i}{dt^2} \quad (3.3)$$

Here, $\phi_i(t)$ is the superconducting phase difference across the $i$th junction, $\Phi_0$ is the magnetic flux quantum, and $I_C$ is the critical current of each junction. In the case of our model with four junctions, the fluxoid quantization condition, which relates each $\phi_i(t)$ to the external flux, reads:

$$\phi_i = \frac{\pi}{2} n + \frac{\pi}{2} \frac{\Phi_{TOT}}{\Phi_0} \quad (3.4)$$

where $n$ is an integer and, by symmetry, we assume that (Araujo-Moreira *et al* 1997, Barbara *et al* 1999) :

$$\phi_1 = \phi_2 = \phi_3 = \phi_4 \equiv \phi_i \quad (3.5)$$

In the case of an oscillatory external magnetic field of the form of Eq. (3.1), the magnetization is given by:

$$M = \frac{LI}{\mu_0 a^2} \quad (3.6)$$

It may be expanded as a Fourier series in the form:

$$M(t) = h_{AC} \sum_{n=0}^{\infty} [\chi_n' \cos(n\omega t) + \chi_n'' \sin(n\omega t)] \quad (3.7)$$

We calculated χ' and χ" through this equation. Both Euler and fourth-order Runge-Kutta integration methods provided the same numerical results. In our model we do not include other effects (such as thermal activation) beyond the above equations. In this case, the temperature-dependent parameter is the critical current of the junctions, given to good approximation by (Meservey 1969, Sergeenkov *et al* 2007):

$$I_C(T) = I_C(0)\left[\frac{\Delta(T)}{\Delta(0)}\right]\tanh\left[\frac{\Delta(T)}{2k_B T}\right] \qquad (3.8)$$

where

$$\Delta(T) = \Delta(0)\tanh\left(2.2\sqrt{\frac{T_C - T}{T}}\right) \qquad (3.9)$$

is the analytical approximation of the BCS gap parameter with $\Delta(0) = 1.76 k_B T_C$.

We simulated $\chi_1$ as a function of temperature and applied magnetic fields keeping in mind that $\chi_1$ depends on the geometrical parameter $\beta_L$ (which is proportional to the number of flux quanta that can be screened by the maximum critical current in the junctions), and the dissipation parameter $\beta_C$ (which is proportional to the capacitance of the junction)

$$\beta_L(T) = \frac{2\pi L I_C(T)}{\Phi_0} \qquad (3.10)$$

$$\beta_C(T) = \frac{2\pi I_C C_J R_J^2}{\Phi_0} \qquad (3.11)$$

## 4. On the origin of dynamic reentrance

According to the current paradigm, paramagnetic Meissner effect (PME) can be related to the presence of $\pi$-junctions (Braunisch *et al* 1992, Li 2003, Kostic *et al* 1996, Ortiz *et al* 2001, Passos *et al* 2000, Lucht *et al* 1995, Li and D. Dominguez 2000), either resulting from the presence of magnetic impurities in the junction (Bulaevskii *et al* 1977, Kusmartsev 1992) or from unconventional pairing symmetry (Kawamura and Li 1996). Other possible explanations of this phenomenon are based on flux trapping (Chen *et al* 1995) and flux compression effects (Terentiev *et al* 1999). Besides, in the experiments with unshunted 2D-JJA, we have previously reported (Araujo-Moreira *et al* 1997, Barbara *et al* 1999) that PME manifests itself through a dynamic reentrance (DR) of the AC magnetic susceptibility as a function of temperature. These results have been further corroborated by Nielsen *et al* (2000) and De Leo *et al* (2001) who argued that PME can be simply related to magnetic screening in multiply connected superconductors. So, the main question is: which parameters are directly responsible for the presence (or absence) of DR in artificially prepared arrays?

Previously (also within the single plaquette approximation), Barbara *et al* (1999) have briefly discussed the effects of varying $\beta_L$ on the observed dynamic reentrance with the main emphasis on the behavior of 2D-JJA samples with high (and fixed) values of $\beta_C$. However, to our knowledge, up to date no systematic study (either experimental or theoretical) has been done on how the $\beta_C$ value itself affects the reentrance behavior. In this section, by a comparative study of the magnetic properties of shunted and unshunted 2D-JJA, we propose an answer to this open question. Namely, by using experimental and theoretical results, we will demonstrate that only arrays with sufficiently large value of the Stewart-McCumber parameter $\beta_C$ will exhibit the dynamic reentrance behavior.

To measure the complex AC susceptibility in our arrays we used a high-sensitive home-made susceptometer based on the so-called screening method in the reflection configuration (Jeanneret *et al* 1989, Araujo-Moreira *et al* 2002), as shown in previous sections. The experimental system was calibrated by using a high-quality niobium thin film. To experimentally investigate the origin of the reentrance, we have measured $\chi'(T)$ for three sets of shunted and unshunted samples obtained from different makers (Westinghouse and Hypress) under the same conditions of the amplitude of the excitation field $h_{ac}$ (1 mOe < $h_{ac}$ < 10 Oe), external magnetic field $H_{dc}$ (0 < $H_{dc}$ < 500 Oe) parallel to the plane of the sample, and frequency of AC field $\omega = 2\pi f$ (fixed at f = 20 kHz). Unshunted 2D-JJAs are formed by loops of niobium islands linked through Nb-AlO$_x$-Nb Josephson junctions while shunted 2D-JJAs have a molybdenum shunt resistor (with $R_{sh} \approx 2.2\Omega$) short-circuiting each junction (see figure 1). Both shunted and unshunted samples have rectangular geometry and consist of $100 \times 150$ tunnel junctions. The unit cell for both types of arrays has square geometry with lattice spacing $a \approx 46\mu m$ and a single junction area of $5 \times 5\mu m^2$. The critical current density for the junctions forming the arrays is about 600A/cm$^2$ at 4.2 K. Besides, for the unshunted samples $\beta_C(4.2K) \approx 30$ and $\beta_L(4.2K) \approx 30$, while for shunted samples $\beta_C(4.2K) \approx 1$ and $\beta_L(4.2K) \approx 30$ where $\beta_L$ and $\beta_C$ are given by expressions (3.10) and (3.11), respectively. There, $C_j \approx 0.58pF$ is the capacitance, $R_j \approx 10.4\Omega$ the quasi-particle resistance (of unshunted array), and $I_C(4.2K) \approx 150\mu A$ the critical current of the Josephson junction. $\Phi_0$ is the quantum of magnetic flux.

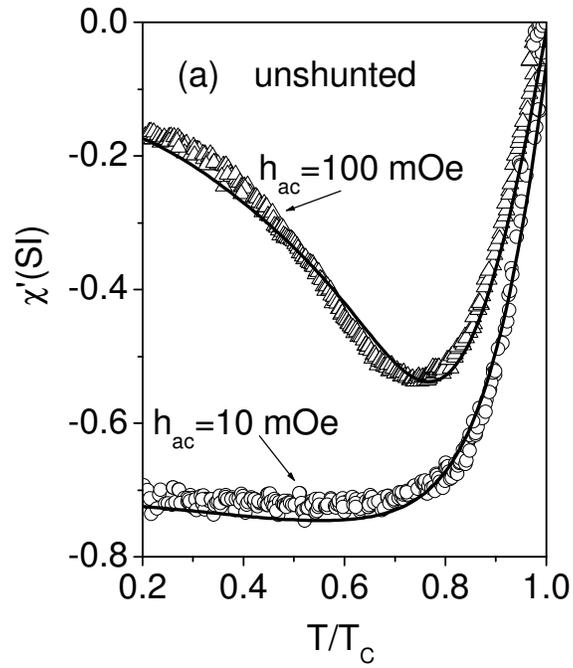

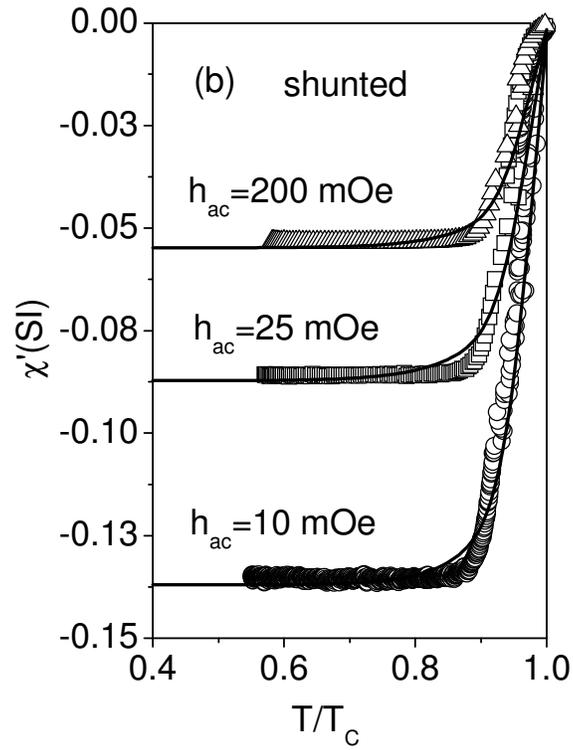

**Figure 5.** Experimental results for $\chi'(T, h_{ac}, H_{dc})$: (a) unshunted 2D-JJA for $h_{ac}$ = 10 and 100 mOe; (b) shunted 2D-JJA for $h_{ac}$ = 10, 25, and 200 mOe. In all these experiments $H_{dc} = 0$. Solid lines are the best fits (see text).

Since our shunted and unshunted samples have the same value of $\beta_L$ and different values of $\beta_C$, it is possible to verify the dependence of the reentrance effect on the value of the Stewart-McCumber parameter. For the unshunted 2D-JJA (figure 5a) we have found that for an AC field lower than 50 mOe the behavior of $\chi'(T)$ is quite similar to homogeneous superconducting samples, while for $h_{ac} > 50$ mOe (when the array is in the mixed-like state with practically homogeneous flux distribution) these samples exhibit a clear reentrant behavior of susceptibility (Araujo-Moreira et al 1999). At the same time, the identical experiments performed on the shunted samples produced no evidence of any reentrance for all values of $h_{ac}$ (see figure 5b). It is important to point out that the analysis of the experimentally obtained imaginary component of susceptibility $\chi''(T)$ shows that for the highest AC magnetic field amplitudes (of about 200 mOe) dissipation remains small. Namely, for typical values of the AC amplitude, $h_{ac} = 100$ mOe (which corresponds to about 10 vortices per unit cell) the imaginary component is about 15 times smaller than its real counterpart. Hence contribution from the dissipation of vortices to the observed phenomena can be safely neglected.

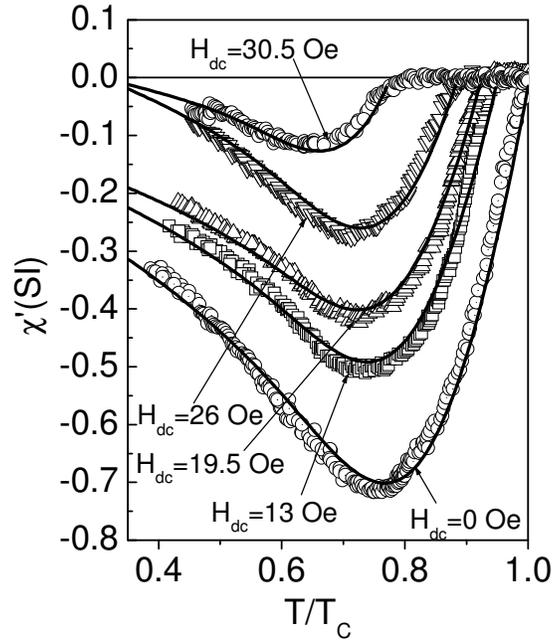

**Figure 6.** Experimental results for $\chi'(T, h_{ac}, H_{dc})$ for unshunted 2D-JJA for $H_{dc} = 0, 13, 19.5, 26,$ and 30.5 Oe. In all these experiments $h_{ac} = 100$ mOe. Solid lines are the best fits (see text).

To further study this unexpected behavior we have also performed experiments where we measure $\chi'(T)$ for different values of $H_{dc}$ keeping the value of $h_{ac}$ constant. The influence of DC fields on reentrance in unshunted samples is shown in figure 6. On the other hand, the shunted samples still show no signs of reentrance, following a familiar pattern of field-induced gradual diminishing of superconducting phase (very similar to a zero DC field flat-like behavior seen in figure 5b).

To understand the influence of DC field on reentrance observed in unshunted arrays, it is important to emphasize that for our sample geometry this parallel field suppresses the critical current $I_C$ of each junction without introducing any detectable flux into the plaquettes of the array. Thus, a parallel DC magnetic field allows us to vary $I_C$ independently from temperature and/or applied perpendicular AC field. The measurements show (see figure 6) that the position of the reentrance is tuned by $H_{dc}$.

We also observe that the value of temperature $T_{min}$ (at which $\chi'(T)$ has a minimum) first shifts towards lower temperatures as we raise $H_{dc}$ (for small DC fields) and then bounces back (for higher values of $H_{dc}$). This non-monotonic behavior is consistent with the weakening of $I_C$ and corresponds to Fraunhofer-like dependence of the Josephson junction critical current on DC magnetic field applied in the plane of the junction. We measured $I_C$ from transport current-voltage characteristics, at different values of $H_{dc}$ at T = 4.2 K and found that $\chi'(T = 4.2K)$, obtained from the isotherm T = 4.2 K (similar to that given in figure 6), shows the same Fraunhofer-like dependence on $H_{dc}$ as the critical current $I_C(H_{dc})$ of the junctions forming the array (see figure 7). This gives further proof that only the junction critical current is varied in this experiment. This also indicates that the screening currents at low temperature (i.e., in the reentrant region) are proportional to the critical currents of the junctions. In addition, this shows an alternative way to obtain $I_C(H_{dc})$ dependence in big arrays. And finally, a sharp Fraunhofer-like pattern observed in both arrays clearly reflects a rather strong coherence (with negligible distribution of critical currents and sizes of the individual junctions) which is based on highly correlated response of *all* single junctions forming the arrays, thus proving their high quality. Such a unique behavior of Josephson junctions in our samples provides a necessary justification for suggested theoretical interpretation of the obtained experimental results. Namely, based on the above-mentioned properties of our arrays, we have found that practically all the experimental results can be explained by analyzing the dynamics of just a single unit cell in the array.

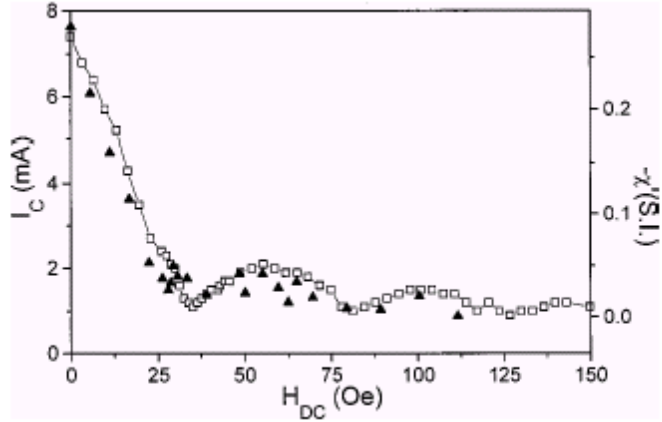

**Figure 7.** The critical current $I_C$ (open squares) and the real part of AC susceptibility $\chi'$ (solid triangles) as a function of DC field $H_{dc}$ for T=4.2K (Araujo-Moreira *et al* 1999).

To understand the different behavior of the AC susceptibility observed in shunted and unshunted 2D-JJAs, in principle one would need to analyze in detail the flux dynamics in these arrays. However, as we have previously reported (Araujo-Moreira *et al* 1999), because of the well-defined periodic structure of our arrays (with no visible distribution of junction sizes and critical currents), it is reasonable to expect that the experimental results obtained from the magnetic properties of our 2D-JJAs can be quite satisfactory explained by analyzing the dynamics of a single unit cell (plaquette) of the array. An excellent agreement between a single-loop approximation and the observed behavior (seen through the data fits) justifies *a posteriori* our assumption. It is important to mention that the idea to use a single unit cell to qualitatively understand PME was first suggested by Auletta *et al* (1994, 1995). They simulated the field-cooled DC magnetic susceptibility of a single-junction loop and found a paramagnetic signal at low values of external magnetic field.

In our calculations and numerical simulations, the unit cell is a loop containing four identical Josephson junctions and the measurements correspond to the zero-field cooling (ZFC) AC magnetic susceptibility. We consider the junctions of the single unit cell as having capacitance $C_j$, quasi-particle resistance $R_j$ and critical current $I_C$. As shown in previous sections, here we have also used this simple four-junctions model to study the magnetic behavior of our 2D-JJA by calculating the AC complex magnetic susceptibility $\chi = \chi' + i\chi''$ as a temperature dependent functional of $\beta_L$ and $\beta_C$. Specifically, shunted samples are identified through low values of the McCumber parameter $\beta_C \approx 1$ while high values $\beta_C \gg 1$ indicate an unshunted 2D-JJA.

If we apply an AC external field $B_{ac}(t) = \mu_0 h_{ac} \cos(\omega t)$ normally to the 2D-JJA and a DC field $B_{dc} = \mu_0 H_{dc}$ parallel to the array, then the total magnetic flux $\Phi(t)$ threading the four-junction superconducting loop is given by $\Phi(t) = \Phi_{ext}(t) + LI(t)$ where L is the loop inductance, $\Phi_{ext}(t) = SB_{ac}(t) + (ld)B_{dc}$ is the flux related to the applied magnetic field (with $l \times d$ being the size of the single junction area, and $S \approx a^2$ being the projected area of the loop), and the circulating current in the loop is described by Eq.(3.3).

Since the inductance of each loop is $L = \mu_0 a \approx 64$ pH, and the critical current of each junction is $I_C \approx 150 \mu A$, for the mixed-state region (above 50 mOe) we can safely neglect the self-field effects because in this region $LI(t)$ is always smaller than $\Phi_{ext}(t)$. Besides, since the length l and the width w of each junction in our array is smaller than the Josephson penetration depth $\lambda_j = \sqrt{\Phi_0 / 2\pi\mu_0 d j_{c0}}$ (where $j_{c0}$ is the critical current density of the junction, and $d = 2\lambda_L + \xi$ is the size of the contact area with $\lambda_L(T)$ being the London penetration depth of the junction and $\xi$ an insulator thickness), namely $l \approx w \approx 5\mu m$ and $\lambda_j \approx 20\mu m$ (using $j_{c0} \approx 600 A/cm^2$ and $\lambda_L \approx 39nm$ for Nb at T = 4.2 K), we can adopt the small-junction approximation (Orlando and Delin 1991) for the gauge-invariant superconducting phase difference across the $i^{th}$ junction

$$\phi_i(t) = \phi_0(H_{dc}) + \frac{2\pi B_{ac}(t)S}{\Phi_0} \qquad (4.1)$$

where $\phi_0(H_{dc}) = \phi_0(0) + 2\pi\mu_0 H_{dc} ld/\Phi_0$ with $\phi_0(0)$ being the initial phase difference.

To properly treat the magnetic properties of the system, let us introduce the following Hamiltonian:

$$H(t) = J\sum_{i=1}^{4}[1-\cos\phi_i(t)] + \frac{1}{2}LI(t)^2 \qquad (4.2)$$

which describes the tunneling (first term) and inductive (second term) contributions to the total energy of a single plaquette. Here, $J(T) = (\Phi_0/2\pi)I_C(T)$ is the Josephson coupling energy.

The real part of the complex AC susceptibility is defined as:

$$\chi'(T, h_{ac}, H_{dc}) = \frac{\partial M}{\partial h_{ac}} \qquad (4.3)$$

where:

$$M(T, h_{ac}, H_{dc}) = -\frac{1}{V}\left\langle \frac{\partial H}{\partial h_{ac}} \right\rangle \qquad (4.4)$$

is the net magnetization of the plaquette. Here V is the sample's volume, and <...> denotes the time averaging over the period $2\pi/\omega$, namely:

$$\langle A \rangle = \frac{1}{2\pi}\int_0^{2\pi} d(\omega t) A(t) \qquad (4.5)$$

Taking into account the analytical approximation of the BCS gap parameter (valid for all temperatures) and for the explicit temperature dependence of the Josephson critical current, given by Eqs.(3.8) and (3.9), we successfully fitted all our data using the following set of parameters: $\phi_0(0) = \pi/2$ (which corresponds to the maximum Josephson current within a plaquette), $\beta_L(0) = \beta_C(0) = 32$ (for unshunted array) and $\beta_C(0) = 1.2$ (for shunted array). The corresponding fits are shown by solid lines in figures 5 and 6 for the experimental values of AC and DC field amplitudes.

In the mixed-state region and for low enough frequencies (this assumption is well-satisfied because in our case $\omega \ll \omega_{LR}$ and $\omega \ll \omega_{LC}$ where $\omega_{LR} = R/L$ and $\omega_{LC} = 1/\sqrt{LC}$ are the two characteristic frequencies of the problem) from Eqs.(4.3)-(4.5) we obtain the following approximate analytical expression for the susceptibility of the plaquette:

$$\chi'(T, h_{ac}, H_{dc}) \approx -\chi_0(T)\left[\beta_L(T)f_1(b)\cos\left(\frac{2H_{dc}}{H_0}\right) + f_2(b)\sin\left(\frac{H_{dc}}{H_0}\right) - \beta_C^{-1}(T)\right] \quad (4.6)$$

where $\chi_0(T) = \pi S^2 I_C(T)/V\Phi_0$, $H_0 = \Phi_0/(2\pi\mu_0 dl) \approx 10$ Oe, $f_1(b) = J_0(2b) - J_2(2b)$, $f_2(b) = J_0(b) - bJ_1(b) - 3J_2(b) + bJ_3(b)$ with $b = 2\pi S\mu_0 h_{ac}/\Phi_0$ and $J_n(x)$ being the Bessel function of the $n^{th}$ order.

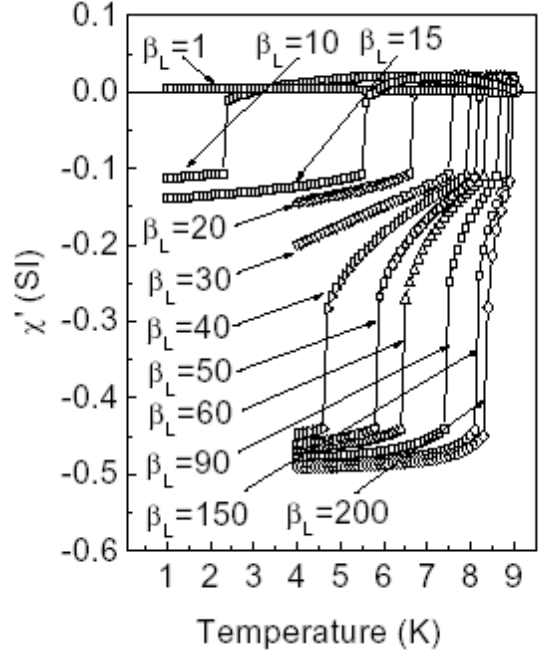

**Figure 8.** Numerical simulation results for $h_{ac} = 70$ mOe, $H_{dc} = 0$, $\beta_C(T=4.2K) = 1$ and for different values of $\beta_L(T=4.2K)$ based on Eqs.(4.4)-(4.6).

Notice also that the analysis of Eq.(4.6) reproduces the observed Fraunhofer-like behavior of the susceptibility in applied DC field (figure 7) and the above-mentioned fine tuning of the reentrance effect. Indeed, according to Eq.(4.6) (and in agreement with the observations), for small DC fields the temperature $T_{min}$ (indicating the beginning of the reentrant transition) varies with $H_{dc}$ as follows, $(T_C - T_{min})/T_C \approx H_{dc}/H_0$.

To further test our interpretation and verify the influence of the parameter $\beta_C$ on the reentrance, we have also performed extensive numerical simulations of the four-junction model previously described but without a simplifying assumption about the explicit form of the phase difference based on Eq.(4.1). More precisely, we obtained the temperature behavior of the susceptibility by solving the set of equations responsible for the flux dynamics within a single plaquette and based on Eq.(3.3) for the total current $I(t)$, the equation for the total flux $\Phi(t) = \Phi_{ext}(t) + LI(t)$ and the flux quantization condition for four junctions, namely $\phi_i(t) = (\pi/2)[n + (\Phi/\Phi_0)]$ where n is an integer. Both Euler and fourth-order Runge-Kutta integration methods provided the same numerical results. In figure 8 we show the real component of the simulated susceptibility $\chi(T)$ corresponding to the fixed value of $\beta_C(T = 4.2K) = 1$ (shunted samples) and different values of $\beta_L(T = 4.2K)$. As expected, for this low value of $\beta_C$ reentrance is not observed for any values of $\beta_L$. On the other hand, figure 9 shows the real component of the simulated $\chi(T)$ but now using fixed value of $\beta_L(T = 4.2K) = 30$ and different values of $\beta_C(T = 4.2K)$. This figure clearly shows that reentrance appears for values of $\beta_C > 20$. In both cases we used $h_{ac}$=70 mOe. We have also simulated the curve for shunted ($\beta_L = 30$, $\beta_C = 1$) and unshunted ($\beta_L = 30$, $\beta_C = 30$) samples for different values of $h_{ac}$ (see figure 10). In this case the values of the parameters $\beta_L$ and $\beta_C$ were chosen from our real 2D-JJA samples. Again, our simulations confirm that dynamic reentrance does not occur for low values of $\beta_C$, independently of the values of $\beta_L$ and $h_{ac}$.

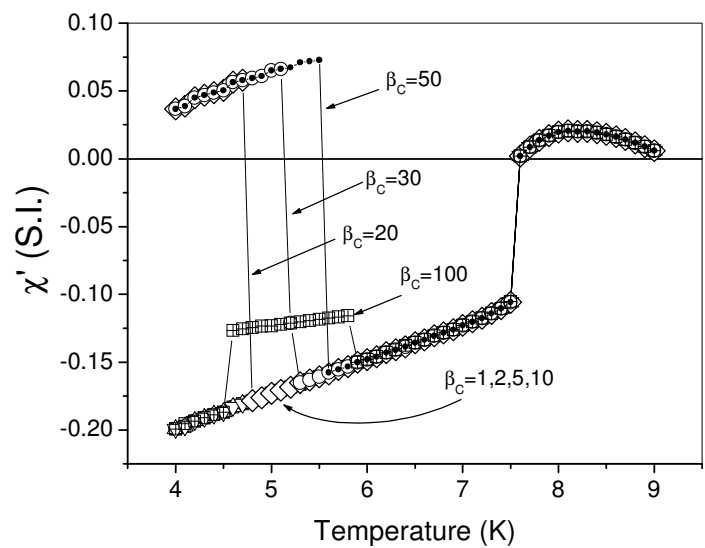

**Figure 9.** Numerical simulation results for $h_{ac} = 70$ mOe, $H_{dc} = 0$, $\beta_L(T = 4.2K) = 30$ and for different values of $\beta_C(T = 4.2K)$ based on Eqs.(4.4)-(4.6).

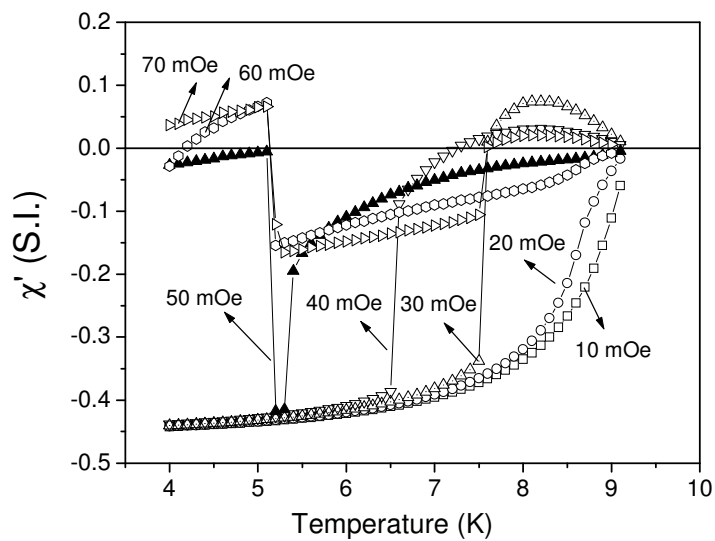

(a)

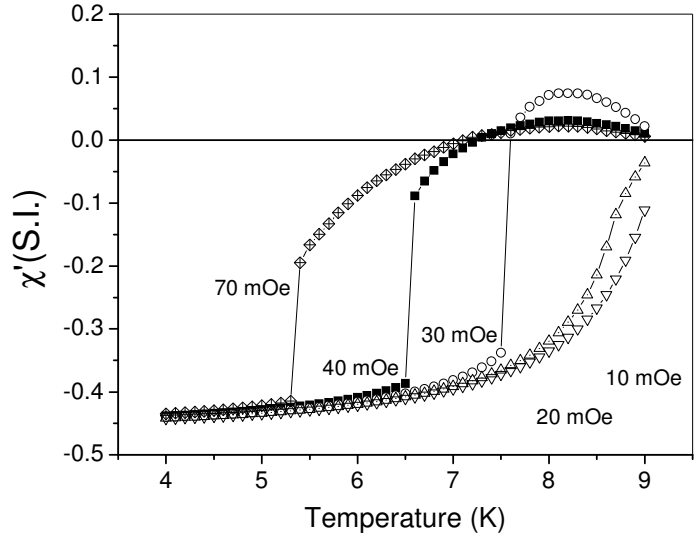

(b)

**Figure 10.** Curves of the simulated susceptibility ($H_{dc} = 0$ and for different values of $h_{ac}$) corresponding to (a) unshunted 2D-JJA with $\beta_L(T = 4.2K) = 30$ and $\beta_C(T = 4.2K) = 30$; (b) shunted 2D-JJA with $\beta_L(T = 4.2K) = 30$ and $\beta_C(T = 4.2K) = 1$.

Based on the above extensive numerical simulations, a resulting *phase diagram* $\beta_C$-$\beta_L$ (taken for T=1K, $h_{ac}$=70 mOe, and $H_{dc}$=0) is depicted in figure 11 which clearly demarcates the border between the reentrant (white area) and non-reentrant (shaded area) behavior in the arrays for different values of $\beta_L(T)$ and $\beta_C(T)$ parameters at given temperature. In other words, if $\beta_L$ and $\beta_C$ parameters of any realistic array have the values inside the white area, this array will exhibit a reentrant behavior.

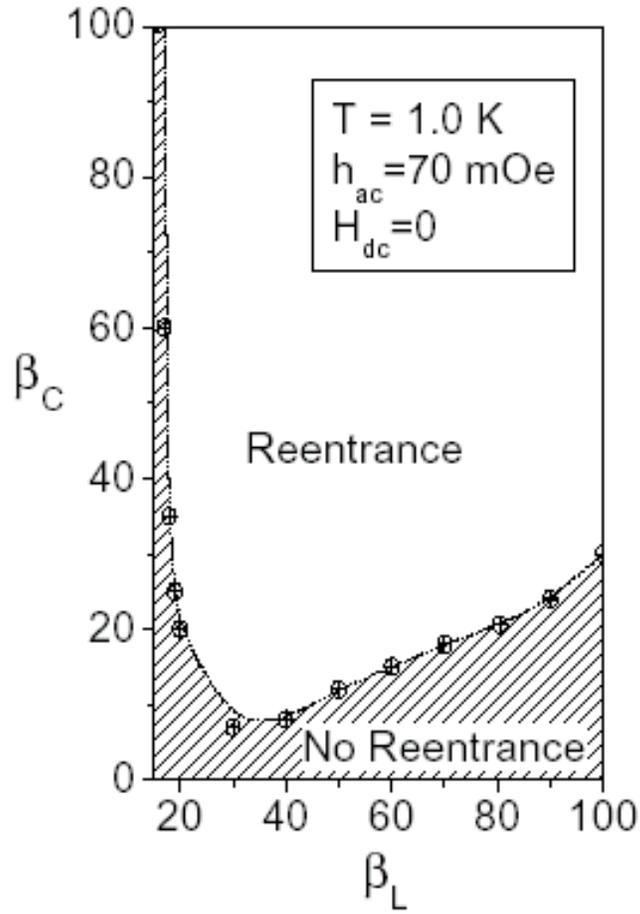

**Figure 11.** Numerically obtained phase diagram (taken for $T = 1K$, $h_{ac}$ = 70 mOe, and $H_{dc} = 0$) which shows the border between the reentrant (white area) and non-reentrant (shaded area) behavior in the arrays for different values of $\beta_L$ and $\beta_C$ parameters.

It is instructive to mention that a hyperbolic-like character of $\beta_L$ vs. $\beta_C$ law (seen in figure 11) is virtually present in the approximate analytical expression for the susceptibility of the plaquette given by Eq.(4.6) (notice however that this expression can not be used to produce any quantitative prediction because the neglected in Eq.(4.6) frequency-related terms depend on $\beta_L$ and $\beta_C$ parameters as well). A qualitative behavior of the envelope of the phase diagram (depicted in figure 11) with DC magnetic field $H_{dc}$ (for T=1 K and $h_{ac}$=70 mOe), obtained using Eq.(4.6), is shown in figure 12.

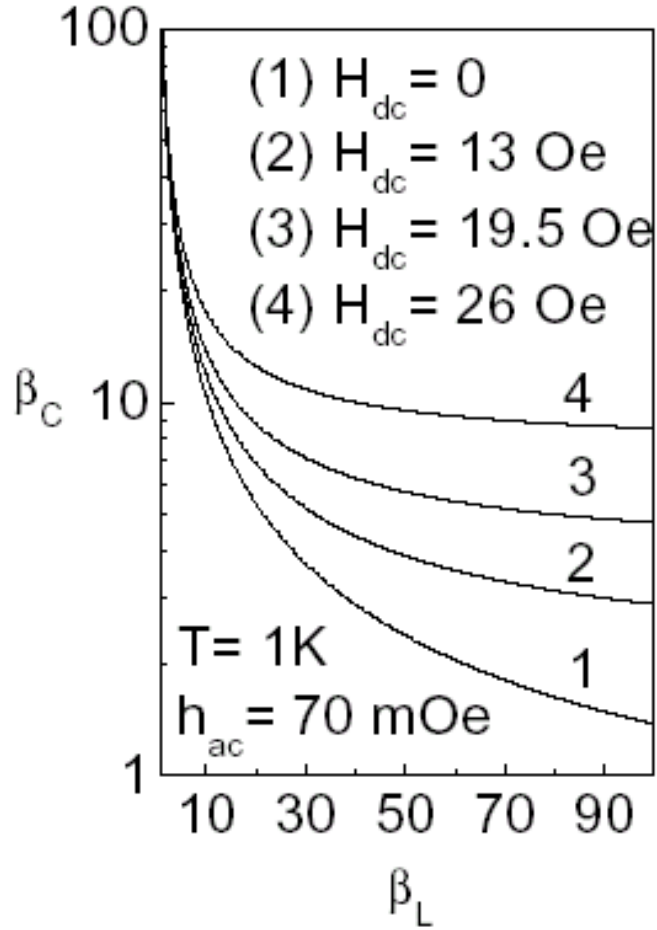

**Figure 12.** A qualitative behavior of the envelope of the phase diagram (shown in previous figure) with DC magnetic field $H_{dc}$ (for $T = 1K$ and $h_{ac} = 70mOe$) obtained from Eq.(4.6).

And finally, to understand how small values of $\beta_C$ parameter affect the flux dynamics in shunted arrays, we have analyzed the $\Phi_{tot} - \Phi_{ext}$ diagram. Similarly to those results previously obtained from unshunted samples (Araujo-Moreira *et al* 1999), for a shunted sample at fixed temperature this curve is also very hysteretic (see figure 13). In both cases, $\Phi_{tot}$ vs. $\Phi_{ext}$ shows multiple branches intersecting the line $\Phi_{tot} = 0$ which corresponds to diamagnetic states. For all the other branches, the intersection with the line $\Phi_{tot} = \Phi_{ext}$ corresponds to the boundary between diamagnetic states (negative values of $\chi'$) and paramagnetic states (positive values of $\chi'$). As we have reported before (Araujo-Moreira *et al* 1999), for unshunted 2D-JJA at temperatures below 7.6 K the appearance of the first and third branches adds a paramagnetic contribution to the average value of $\chi'$. When $\beta_C$ is small (shunted arrays), the analysis of these curves shows that there is no reentrance at low temperatures because in this case the second branch appears to be energetically stable, giving an extra diamagnetic contribution which overwhelms the paramagnetic contribution from subsequent branches. In other words, for low enough values of $\beta_C$ (when the samples are ZFC and then measured at small values of the magnetic field), most of the loops will be in the diamagnetic states, and no paramagnetic response is registered. As a result, the flux quanta cannot get trapped into the loops even by the following field-cooling process in small values of the magnetic field. In this case the superconducting phases and the junctions will have the same diamagnetic response and the resulting measured value of the magnetic susceptibility will be negative (i.e., diamagnetic) as well. On the other hand, when $\beta_C$ is large enough (unshunted arrays), the second branch becomes energetically unstable, and the average response of the sample at low temperatures is paramagnetic (Araujo-Moreira *et al* 1999).

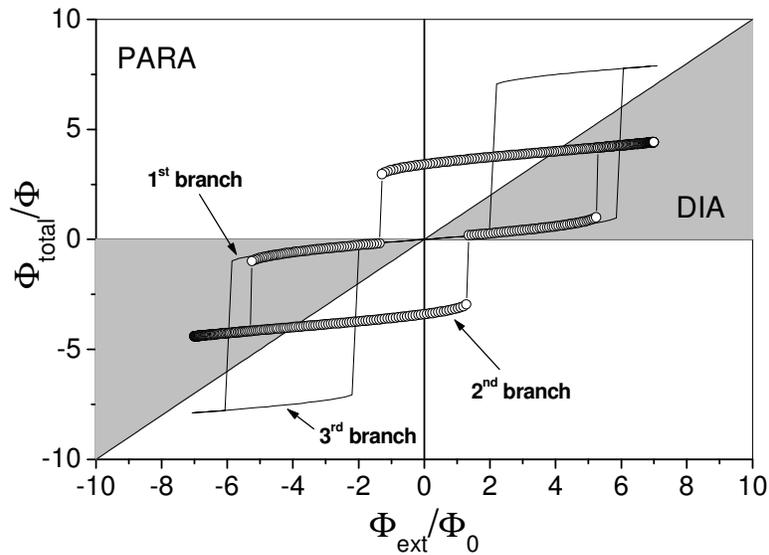

**Figure 13.** Numerical simulation results, based on Eqs.(4.3)-(4.6), showing $\Phi_{tot}$ vs. $\Phi_{ext}$ for shunted 2D-JJA with $\beta_L(T=4.2K)=30$ and $\beta_C(T=4.2K)=1$.

## 5. Manifestation of geometry imposed quantization effects in temperature behavior of AC magnetic response.

Among the numerous spectacular phenomena recently discussed and observed in 2D-JJAs probably one of the most fascinating and intriguing is the so-called avalanche-like magnetic field behavior of magnetization (Ishikaev *et al* 2000, Altshuler and Johansen 2004) which is closely linked to self-organized criticality (SOC) phenomenon (Jensen 1998, Ginzburg and Savitskaya 2001). More specifically, using highly sensitive SQUID magnetometer, magnetic field jumps in the magnetization curves associated with the entry and exit of avalanches of tens and hundreds of fluxons were clearly seen in SIS-type arrays (Ishikaev *et al* 2000). Besides, it was shown that the probability distribution of these processes is in good agreement with the SOC theory (Ginzburg and Savitskaya 2001). An avalanche character of flux motion was observed at temperatures at which the size of the fluxons did not exceed the size of the cell, that is, for discrete vortices. On the other hand, using a similar technique, magnetic flux avalanches were not observed in SNS-type proximity arrays (Ishikaev *et al* 2002) despite a sufficiently high value of the inductance L related critical parameter $\beta_L = 2\pi L I_C / \Phi_0$ needed to satisfy the observability conditions of SOC. Instead, the observed quasi-hydrodynamic flux motion in the array was explained by the considerable viscosity characterizing the vortex motion through the Josephson junctions.

In this section we present experimental evidence for manifestation of novel geometric effects in magnetic response of high-quality ordered 2D-JJA (Sergeenkov and Araujo-Moreira 2004). By improving resolution of home-made mutual-inductance measurements technique described in the beginning of this article, a pronounced step-like structure (with the number of steps n = 4 for all AC fields) has been observed in the temperature dependence of AC susceptibility in artificially prepared two-dimensional Josephson Junction Arrays (2D-JJA) of unshunted Nb-AlO$_x$-Nb junctions with $\beta_L(4.2K) = 30$. Using a single-plaquette approximation of the overdamped 2D-JJA model, we were able to successfully fit our data assuming that steps are related to the geometric properties of the plaquette. The number of steps n corresponds to the number of flux quanta that can be screened by the maximum critical current of the junctions. The steps are predicted to manifest themselves in arrays with the inductance related parameter $\beta_L$ matching a "quantization" condition $\beta_L(0) = 2\pi(n+1)$.

To measure the complex AC susceptibility in our arrays with high precision, we used a home-made susceptometer based on the so-called screening method in the reflection configuration as described in the previous sections. Measurements were performed as a function of the temperature T (for 1.5 K < T < 15 K), and the amplitude of the excitation field $h_{ac}$ (for 1 mOe < $h_{ac}$ < 10 Oe) normal to the plane of the array. The frequency of AC field in the experiments reported here was fixed at 20 kHz. The used in the present study unshunted 2D-JJAs are formed by loops of niobium islands (with $T_C$ = 9.25 K) linked through Nb-AlO$_x$-Nb Josephson junctions and consist of 100×150 tunnel junctions described in previous sections.

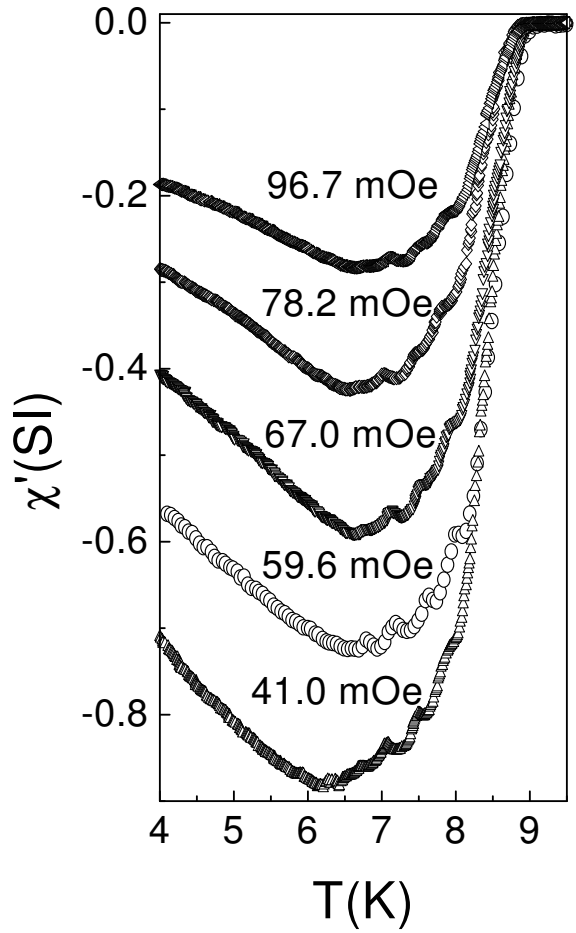

(a)

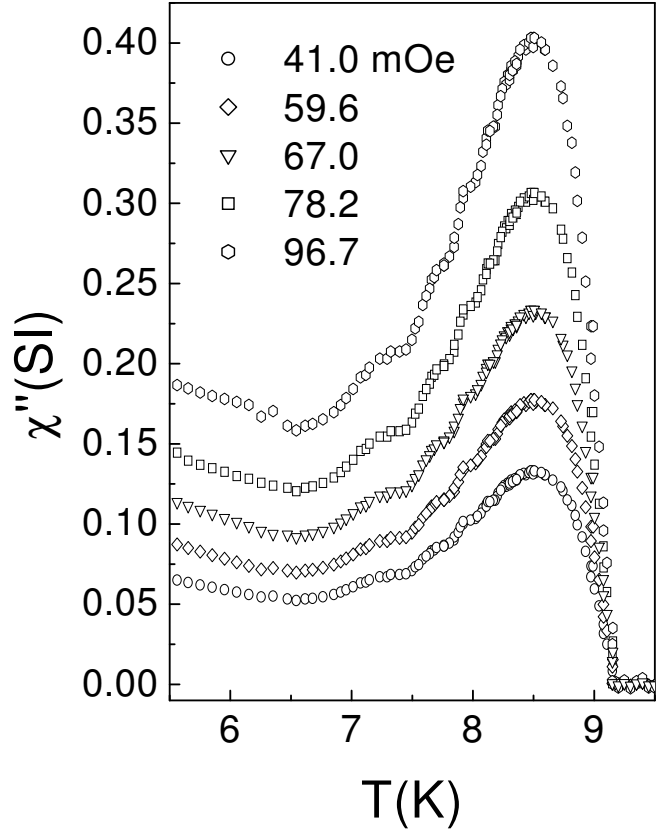

**(b)**

**Figure 14.** Experimental results for temperature dependence of the real (a) and imaginary (b) parts of AC susceptibility $\chi'(T, h_{ac})$ for different AC field amplitudes $h_{ac}$ = 41.0, 59.6, 67.0, 78.2 and 96.7 mOe.

It is important to recall that the magnetic field behavior of the critical current of the array (taken at T=4.2 K) on DC magnetic field $H_{dc}$ (parallel to the plane of the sample) exhibited a sharp Fraunhofer-like pattern characteristic of a single-junction response, thus proving a rather strong coherence within arrays (with negligible distribution of critical currents and sizes of the individual junctions) and hence the high quality of our sample.

The observed temperature dependence of both the real and imaginary parts of AC susceptibility for different AC fields is shown in figure 14. A pronounced step-like structure is clearly seen at higher temperatures. The number of steps n does not depend on AC field amplitude and is equal to n = 4. As expected (Araujo-Moreira *et al* 2005), for $h_{ac}$ > 40 mOe (when the array is in the mixed-like state with practically homogeneous flux distribution) the steps are accompanied by the previously observed reentrant behavior with $\chi'(T, h_{ac})$ starting to increase at low temperatures.

To understand the step-like behavior of the AC susceptibility observed in unshunted 2D-JJAs, in principle one would need to analyze in detail the flux dynamics in these arrays. However, as we have previously reported (Araujo-Moreira *et al* 1999, 2005) because of the well-defined periodic structure of our arrays with no visible distribution of junction sizes and critical currents, it is quite reasonable to assume that the experimental results obtained from the magnetic properties of our 2D-JJAs could be understood by analyzing the dynamics of just a single unit cell (plaquette) of the array. As we shall see, theoretical interpretation of the presented here experimental results based on single-loop approximation, is in excellent agreement with the observed behavior. In our analytical calculations, the unit cell is the loop containing four identical Josephson junctions described in previous sections, and the measurements correspond to the zero-field cooling AC magnetic susceptibility. If we apply an AC external field $H_{ac}(t) = h_{ac} \cos \omega t$ normally to the 2D-JJA, then the total magnetic flux $\Phi(t)$ threading the four-junction superconducting loop is given again by $\Phi(t) = \Phi_{ext}(t) + LI(t)$ where L is the loop inductance, $\Phi_{ext}(t) = SH_{ac}(t)$ is the flux related to the applied magnetic field (with $S \approx a^2$ being the projected area of the loop), and the circulating current in the loop reads $I(t) = I_C(T)\sin\phi(t)$. Here $\phi(t)$ is the gauge-invariant superconducting phase difference across the i$^{th}$ junction. As is well-known, in the case of four junctions, the flux quantization condition reads (Barone and Paterno 1982)

$$\phi = \frac{\pi}{2}\left(n + \frac{\Phi}{\Phi_0}\right) \qquad (5.1)$$

where *n* is an integer, and, for simplicity, we assume as usual that $\phi_1 = \phi_2 = \phi_3 = \phi_4 \equiv \phi$. To properly treat the magnetic properties of the system, let us introduce the following Hamiltonian

$$H(t) = J(T)[1 - \cos\phi(t)] + \frac{1}{2}LI^2(t) \qquad (5.2)$$

which describes the tunneling (first term) and inductive (second term) contributions to the total energy of a single plaquette. Here, $J(T) = (\Phi_0/2\pi)I_C(T)$ is the Josephson coupling energy.

Since the origin of reentrant behavior in our unshunted arrays has been discussed in detail in the previous section of this paper, in what follows we concentrate only on interpretation of the observed here step-like structure of $\chi'(T, h_{ac})$. First of all, we notice that the number of observed steps n (in our case n = 4) clearly hints at a possible connection between the observed here phenomenon and flux quantization condition within a single four-junction plaquette. Indeed, the circulating in the loop current $I(t) = I_C(T)\sin\phi(t)$ passes through its maximum value whenever $\phi(t)$ reaches the value of $(\pi/2)(2n+1)$ with n = 0,1,2... As a result, the maximum number of fluxons threading a single plaquette (see Eq. (5.1)) over the period $2\pi/\omega$ becomes equal to $<\Phi(t)> = (n+1)\Phi_0$. In turn, the latter equation is equivalent to the following condition $\beta_L(T) = 2\pi(n+1)$. Since this formula is valid for any temperature, we can rewrite it as a geometrical "quantization" condition $\beta_L(0) = 2\pi(n+1)$. Recall that in the present experiment, our array has $\beta_L(0) = 31.6$ (extrapolated from its experimental value $\beta_L(4.2K) = 30$) which is a perfect match for the above "quantization" condition predicting n = 4 for the number of steps in a single plaquette, in excellent agreement with the observations.

Based on the above discussion, we conclude that in order to reproduce the observed temperature steps in the behavior of AC susceptibility, we need a particular solution to Eq.(5.1) for the phase difference in the form of $\phi_n(t) = (\pi/2)(2n+1) + \delta\phi(t)$ assuming $\delta\phi(t) \ll 1$. After substituting this Ansatz into Eq.(5.1), we find that $\phi_n(t) \approx (\pi/2)n + (1/4)\beta_L(T) + (1/4)f\cos(\omega t)$ where $f = 2\pi Sh_{ac}/\Phi_0$ is the AC field related frustration parameter. Using this effective phase difference, we can calculate the AC response of a single plaquette. Namely, the real and imaginary parts of susceptibility read

$$\chi'(T, h_{ac}) = \frac{1}{\pi}\int_0^\pi d(\omega t)\cos(\omega t)\chi_n(t) \qquad (5.3a)$$

$$\chi''(T,h_{ac}) = \frac{1}{\pi}\int_0^\pi d(\omega t)\sin(\omega t)\chi_n(t) \quad (5.3b)$$

where

$$\chi_n(t) = -\frac{1}{V}\left[\frac{\partial^2 H}{\partial h_{ac}^2}\right]_{\phi=\phi_n(t)} \quad (5.4)$$

Here V is the sample's volume.

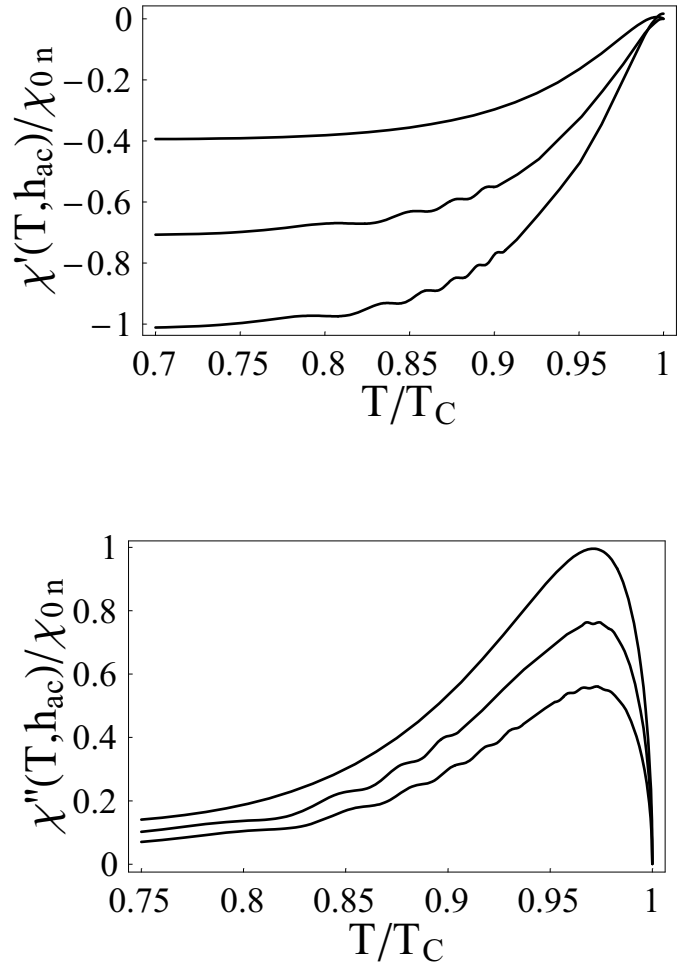

**Figure 15.** Theoretically predicted dependence of the real (upper) and imaginary (lower) parts of normalized susceptibility on reduced temperature according to Eqs.(5.3)-(5.5) for f=0.5 and for "quantized" values of $\beta_L(0) = 2\pi(n+1)$ (from top to bottom): n=0, 3 and 5.

As in the previous sections of this paper, for the explicit temperature dependence of $\beta_L(T) = 2\pi L I_C(T)/\Phi_0$ we again use the analytical approximation of the BCS gap parameter (valid for all temperatures), $\Delta(T) = \Delta(0)\tanh(2.2\sqrt{(T_C - T)/T})$ which governs the temperature dependence of the Josephson critical current:

$$I_C(T) = I_C(0)\left[\frac{\Delta(T)}{\Delta(0)}\right]\tanh\left[\frac{\Delta(T)}{2k_B T}\right] \quad (5.5)$$

Figure 15 depicts the predicted by Eqs.(5.3)-(5.5) dependence of the AC susceptibility on reduced temperature for f=0.5 and for different "quantized" values of $\beta_L(0) = 2\pi(n+1)$. Notice the clear appearance of three and five steps for n = 3 and n = 5, respectively (as expected, n = 0 corresponds to a smooth temperature behavior without steps).

In figure 16 we present fits (shown by solid lines) of the observed temperature dependence of the normalized susceptibility $\chi(T, h_{ac})/\chi_0$ for different magnetic fields $h_{ac}$ according to Eqs.(5.3)-(5.5) using $\beta_L(0) = 10\pi$. As is seen, our simplified model based on a single-plaquette approximation demonstrates an excellent agreement with the observations.

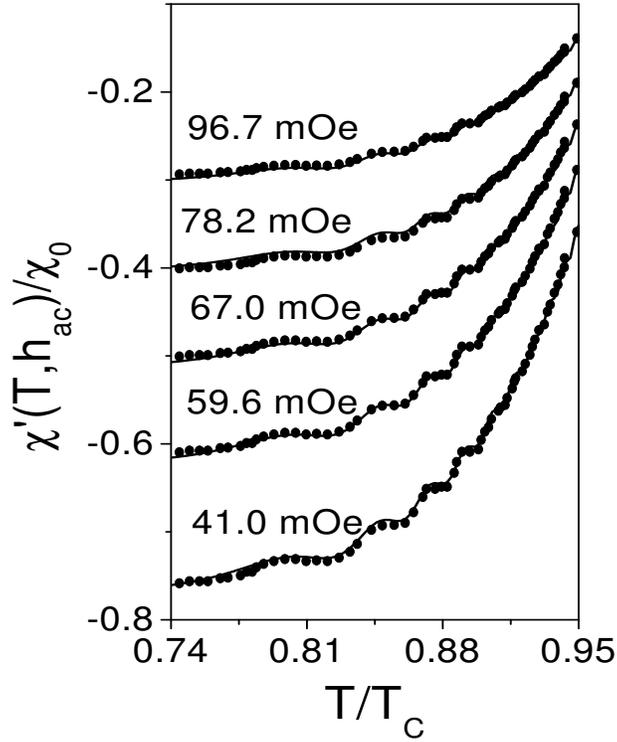

(a)

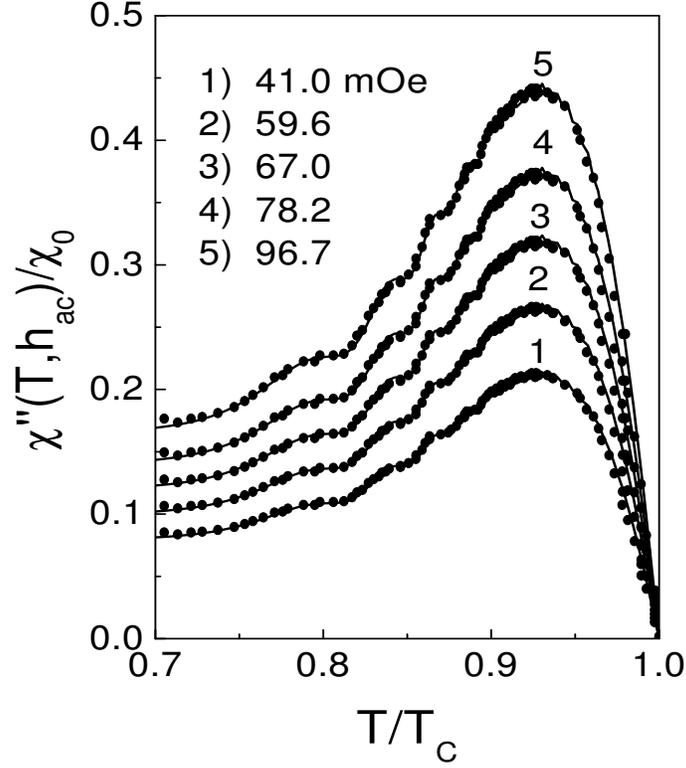

**(b)**

**Figure 16.** Fits (solid lines) of the experimental data for $h_{ac}$ = 41.0, 59.6, 67.0, 78.2, and 96.7 mOe according to Eqs.(5.3)-(5.5) with $\beta_L(0) = 10\pi$.

## 6. Probing distribution of magnetic flux in unshunted JJA via scanning SQUID microscope

To study the distribution of magnetic flux in our unshunted JJA, we have performed several experiments by using a scanning SQUID microscope. The total area of the samples is of about 5.0 mm x 10.0 mm. The analysis of small areas (experiments 1-3) exhibited the typical distribution of magnetic flux expected for a critical state behavior which is not seen on large areas. Moreover, this last result (experiment 4) clearly indicates a dendritic flux distribution. Therefore, a critical state is established only at short range distances, of the order of some rows. We describe below each performed experiment as well as the obtained results.

**Experiment 1**: The image shown in Fig. 17 was taken from a small section of the sample (0.5 mm x 1.8 mm). The sample was zero field cooled (at zero external flux). The scale on the image is normalized to $\Phi/\Phi_0$ in the array, where the x and y axes are in millimeters. This image shows the typical distribution of magnetic flux expected for a critical state behavior. Thus, penetrated field varies almost smoothly from the edge to inside the sample.

**Experiment 2**: Here we have obtained images (shown in Fig. 18) of a small section of the sample (0.5 mm x 1.5 mm) at T = 4.2 K. The sample was ZFC and then an external field was applied normal to the sample. In this case, 1.86 mA in the solenoid coil is equal to 1 flux quantum in the SQUID. The SQUID is 10 x 10 $\mu m^2$. The field was started at 0.0 mA and then ramped up. In the first image, we observe the typical distribution of magnetic flux expected for a critical state. It is possible to observe that the magnetic field inside the sample is increasing for higher applied field.

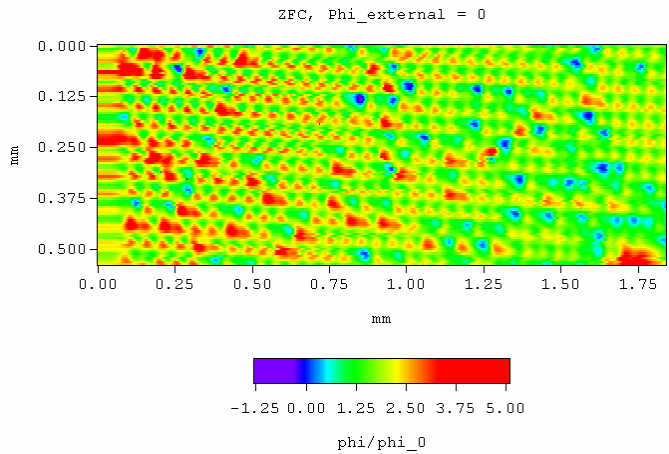

**Figure 17**. SSM images of a small section of the sample after ZFC and for no external magnetic flux. Dimensions of scanned area are shown in millimiters.

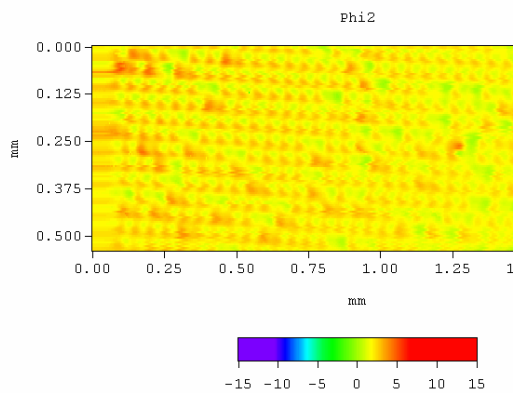

(a) $I_{coil}$ = 0.0 mA

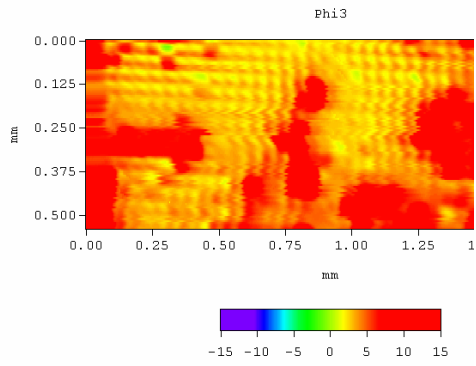

(b) $I_{coil}$ = 0.2 mA

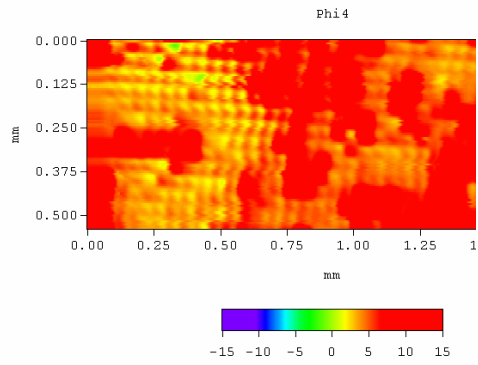

(c) $I_{coil}$ = 0.4 mA

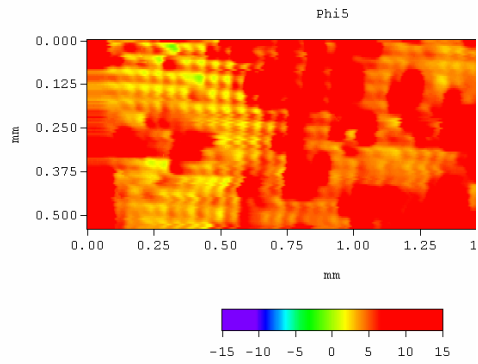

(d) $I_{coil}$ = 0.6 mA

**Figure 18**. SSM images of a small section of the sample at T = 4.2 K, after ZFC. The external magnetic field was started at 0.0 mA and then ramped up and down. Dimensions of scanned area are shown in millimeters.

**Experiment 3**: In this experiment (Fig. 19) we ZFC the sample and then raised the temperature of the sample. The current in the coil (external magnetic field) was kept at 0.2 mA after the ZFC procedure. The scan area is roughly 0.5 mm x 1.75 mm. The color values represent output voltage from the SQUID electronics. Here, 1.72V is equal to 1 flux quantum in the SQUID. These images show the typical distribution of magnetic flux expected for a critical state behavior with penetrated field varying almost smoothly from the edge to inside the sample.

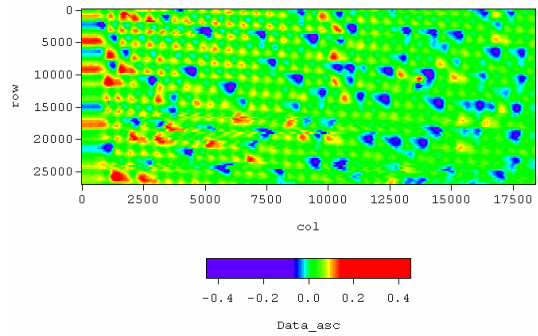

(a) T=4.2 K

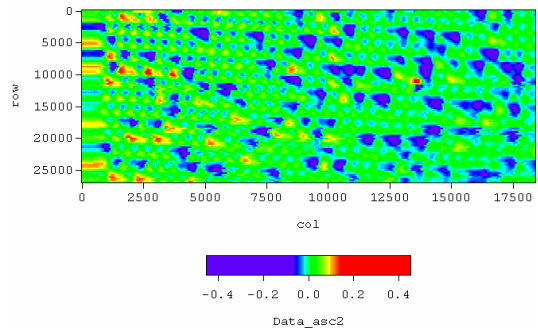

(b) T=5.0 K

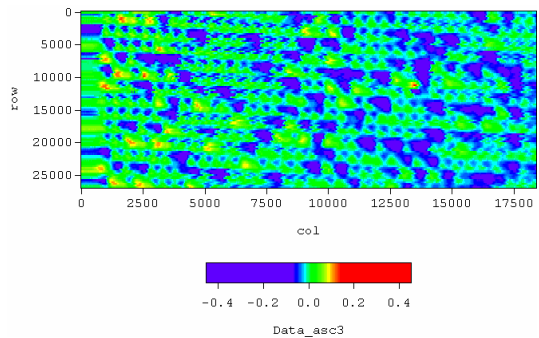

(c) T=5.5 K

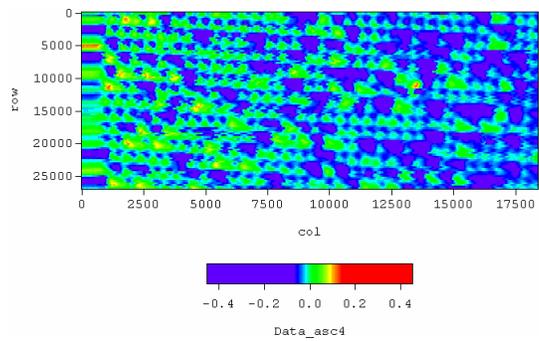

(d) T=6.0 K

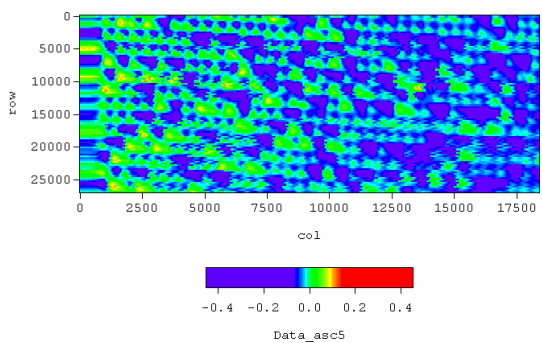

(e)  T=6.5 K

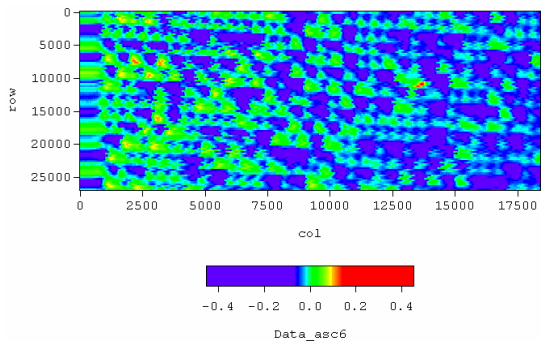

(f) T=7.0 K

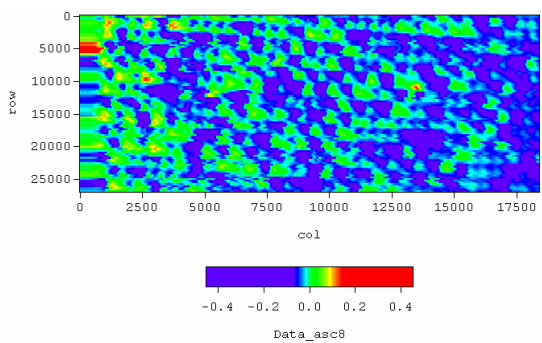

(g) T=7.4 K

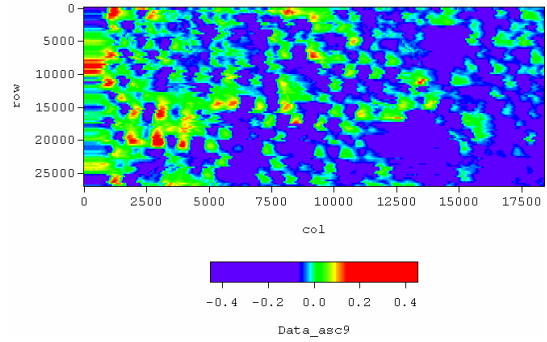

(h) T=7.8 K

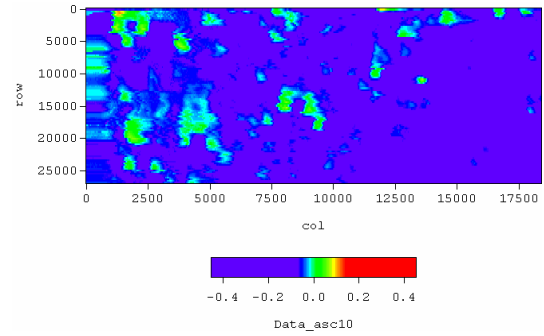

(i) T=8.1 K

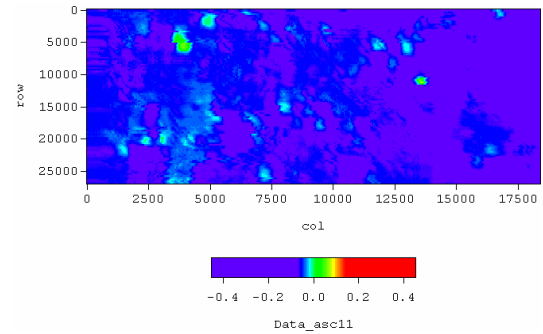

(j) T=8.6 K

**Figure 19**. SSM images of a small section of the sample after ZFC. The images show the evolution of the penetration of magnetic field ($I_{coil}$ = 0.2 mA) as the temperature is raised. Dimensions of scanned area are shown in millimeters.

**Experiment 4**: Images of successive scans (Fig. 20) of a large area (5.0 mm x 3.5 mm) of the sample, for T = 4.2 K, as the external magnetic field is increased. Here we observe that, for long-range distances, the picture of uniform flux fronts, as observed in the preceding images, breaks down. Clearly, the penetration of the magnetic field takes place through the growth of magnetic dendrites. The sudden penetration of magnetic flux at this temperature is consistent with results obtained from AC magnetic susceptibility measurements and from numerical simulations shown in the preceding sections. Thus, through these images we confirm that in 2D-JJA the typical picture of a critical state is valid only in short-range distances. For long distances, the penetration of the field takes place through the growth of magnetic dendrites (Durán *et al* 1995).

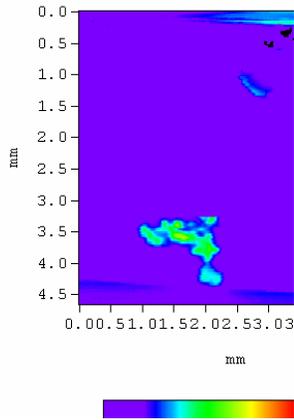 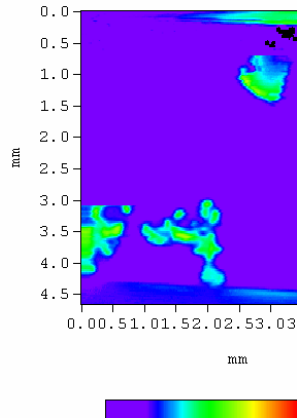

(a) After ZFC  (b) $I_{coil} = 0.05$ mA

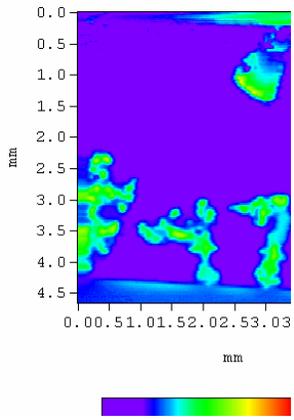 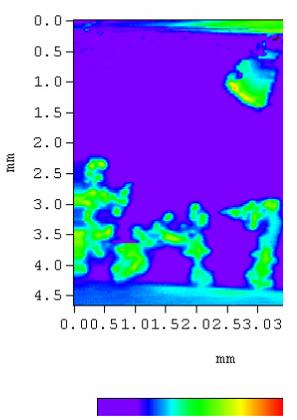

(c) $I_{coil} = 0.10$ mA  (d) $I_{coil} = 0.15$ mA

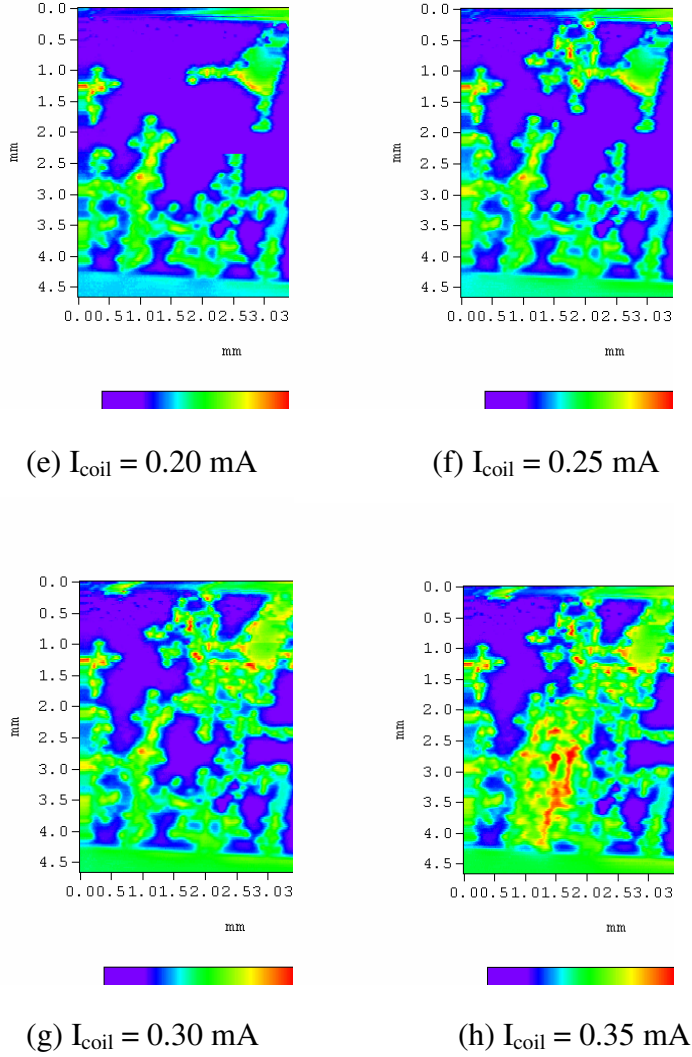

(e) $I_{coil} = 0.20$ mA  (f) $I_{coil} = 0.25$ mA

(g) $I_{coil} = 0.30$ mA  (h) $I_{coil} = 0.35$ mA

**Figure 20**. Large sweep SSM images as the external magnetic field is increased for T = 4.2 K.; x and y axes are in millimeters.

## 7. Conclusions

In this review article three novel interesting phenomena related to the magnetic properties of 2D-JJA were reported. First of all, our experimental and theoretical results have demonstrated that the reentrance of AC susceptibility (and concomitant PME) in artificially prepared 2D-JJA takes place in the underdamped (unshunted) array (with large enough value of the Stewart-McCumber parameter $\beta_C$) and totally disappears in over-damped (shunted) arrays. On the other hand, we have presented a step-like structure (accompanied by previously seen low-temperature reentrance phenomenon) which has been observed in the temperature dependence of AC susceptibility in our artificially prepared 2D-JJA of unshunted Nb-AlO$_x$-Nb junctions. The steps are shown to occur in arrays with the geometry sensitive parameter $\beta_L(T)$ matching the "quantization" condition $\beta_L(0) = 2\pi(n+1)$ where n is the number of steps. And finally, we demonstrated the use of scanning SQUID microscope for imaging the local flux distribution within our unshunted arrays.

**Acknowledgements**

We thank P. Barbara, C.J. Lobb, A. Sanchez and R.S. Newrock for useful discussions. We thank W. Maluf for his help in running some of the experiments. We gratefully acknowledge financial support from Brazilian Agencies FAPESP and CAPES.